\documentclass[12pt]{iopart}
\usepackage{epsfig}
\usepackage{graphicx}
\usepackage{amsfonts}
\usepackage{color}
\usepackage{multirow}

\begin{document}

\title[The scaling of the density of states]{The  scaling of the density 
of states in systems with resonance states}

\author{Federico  M Pont, Omar  Osenda, and Pablo Serra}
\address{Facultad de Matem\'atica, Astronom\'{\i}a y F\'{\i}sica,
Universidad Nacional de C\'ordoba, and IFEG-CONICET, Ciudad Universitaria,
X5016LAE C\'ordoba, Argentina}
\ead{serra@famaf.unc.edu.ar}
\ead{osenda@famaf.unc.edu.ar}
\ead{serra@famaf.unc.edu.ar}

\begin{abstract}
{  Resonance states of a two-electron quantum dot are studied 
using a variational expansion with both real basis-set functions and 
complex scaling methods. We present numerical evidence about the critical
behavior of the density of states in the region where there are resonances.
The critical behavior is signaled by a strong dependence of some features of
the density of states with the basis-set size used to calculate it. The
 resonance energy and lifetime are obtained using the scaling properties of the
density of states}
\end{abstract}
\date{\today}

\pacs{31.15.ac,03.67.Mn,73.22.-f}
\maketitle
\section{Introduction}

Resonance states are slowly decaying scattering states characterized by a large
but finite lifetime \cite{moisereport,reinhardt1996}. There is a host of
established methods that allow the
calculation of the  energy and lifetime associated to the resonance
\cite{moisereport,reinhardt1996,sajeev2008,bylicki2005,dubau1998,kruppa1999,
kar2004} . In many cases of
interest where complex scaling (analytic dilatation) techniques can be applied,
resonances energies and lifetimes  show up as real and imaginary  part,
respectively,  of
isolated complex eigenvalues of the rotated
Hamiltonian \cite{moisereport}. Though conceptually simple, complex
scaling is difficult to implement in molecular systems, for this kind of
systems the analytical continuation is obtained using absorbing potential
methods \cite{sajeev2009}, which can be roughly grouped in two categories: the
absorbing potentials based on negative imaginary potentials (NIPs), and the
potentials derived from the complex scaling based theories (see
\cite{sajeev2009} and References therein).

Despite the wealth of methods available to deal with the calculation of the
resonance energy the subject constantly receives attention and new methods are
proposed. The new proposals arise, mainly, because almost all the methods have
some drawbacks or because new concepts are applied to the problem. 

Recently Pont {\em et al} \cite{pont2010} and Ferr\'on {\em et al}
\cite{ferron2009} have used the Fidelity \cite{zanardi2006} and
the von Neumann entropy \cite{nielsenbook}, respectively, to obtain the real part of the resonance
energy of a two electron quantum dot. Both concepts, borrowed  from the quantum
information theory, provided  accurate methods  for the calculation of 
resonance-state properties. In
particular the method proposed by Pont {\em et al}  \cite{pont2010} 
have some points in common
with the work of Kar and Ho \cite{kar2004}, both methods rest on the
availability of a
monoparametric family of basis sets, where each basis set
corresponds to a particular value of a free parameter.

Kar and Ho \cite{kar2004} calculate a
variational approximation for the problem which provides an approximate
spectrum. With the spectrum they construct the density of states (DOS) of the
problem and, after this step, it is possible to find the complex eigenenergy of
the resonance fitting the DOS. Their method is quite general but
the fit that is necessary to obtain the resonance energy is by no means
unambiguously defined. In contradistinction the methods proposed  by
Pont {\em et al} \cite{pont2010} do not require any fitting, but to produce the resonance
energy for a given problem (with its corresponding parameters) it is necessary
to find an adequate value of the basis-set free parameter by bisection.

In this work we will present a method to find the energy of a
resonance state using the DOS but without resort to any fitting
or bisection. We will study the resonance state that arise in a two electron
quantum dot when its ground state looses stability and enters into the
continuum. In Section~\ref{sec-model} we briefly present the model and the
variational method used to find an approximate spectrum. In
Section~\ref{sec-density} the DOS is obtained and the energy and lifetime of
the resonance state are founded using finite size scaling techniques. Finally,
in Section~\ref{sec-conclusions} we discuss our results and draw some
conclusions.

\section{The model and basic results}
\label{sec-model}

There are many models of quantum dots, with different symmetries and
interactions (see \cite{bylicki2005,ferron2009} and References therein). In this
work we consider a model with spherical symmetry, with
two
electrons interacting via the Coulomb repulsion. The main results should not be
affected by the particular
potential choice as it is already known that  the near  threshold
behavior and other critical
quantities (such as the critical exponents of the energy and other
observables)
are mostly determined by the range of the involved potentials 
\cite{pont_serra_jpa08}.  Therefore
to model the dot potential we use a short-range potential suitable to apply
the complex
scaling method. After this considerations we propose the following
Hamiltonian $H$ for the system

\begin{equation}
\label{hamiltoniano}
H = -\frac{\hbar^2}{2m} \nabla_{{\mathbf r}_1}^2  
-\frac{\hbar^2}{2m} \nabla_{{\mathbf r}_2}^2  + V(r_1)+V(r_2)+ 
\frac{e^2}{\left|{\mathbf r}_2-{\mathbf r}_1\right|} ,
\end{equation}
where $V(r)=-(V_0/r_0^2)\, \exp{(-r/r_0)}$, ${\mathbf r}_i$ the
position operator of electron $i=1,2$;   $r_0$ and $V_0$
determine  the range and depth of the dot potential.
After re-scaling with $r_0$, in atomic units, the Hamiltonian of Eq.
(\ref{hamiltoniano}) can be written as
\begin{equation}
\label{hamil}
H = -\frac{1}{2} \nabla_{{\mathbf r}_1}^2  
-\frac{1}{2} \nabla_{{\mathbf r}_2}^2 -V_0 e^{-r_1}-V_0
e^{-r_2} + 
\frac{\lambda}{\left|{\mathbf r}_2-{\mathbf r}_1\right|} ,
\end{equation}
where $\lambda=r_0$.

We choose the exponential binding potential to take advantage of its analytical
properties. In particular this potential is well behaved and
the energy of the resonance states can be calculated using complex scaling
methods. So, besides its simplicity, the exponential potential allows us to
obtain independently the energy of the resonance state and a check for our
results. The threshold energy, $\varepsilon$, of Hamiltonian Eq.~(\ref{hamil}),
that is, the one-body ground state energy can be calculated
exactly~\cite{galindo} and is given by \begin{equation}
J_{2\sqrt{2\varepsilon}}\left(\sqrt{2V_0}\right)=0,
\end{equation}
where $J_{\nu}(x)$ is the Bessel function.

The discrete spectrum  and the resonance states of the model given by 
Eq.  (\ref{hamil})  can be obtained approximately 
using ${\cal L}^2$
variational functions \cite{bylicki2005}, \cite{kruppa1999}. So, if
$\left|\psi_j(1,2)\right\rangle$ are the exact eigenfunctions of the
Hamiltonian, we look for variational approximations 

\begin{equation}\label{variational-functions}
\left|\psi_j(1,2)\right\rangle \,  \simeq\, 
\left|\Psi_j(1,2)\right\rangle \, =\, 
\sum_{i=1}^M c^{(j)}_{i} \left| \Phi_i 
\right\rangle \, ,\;\; c^{(j)}_{i} = (\mathbf{c}^{(j)})_i 
\;\;;\;\;j=1,\cdots,M \,.
\end{equation}

\noindent where the $\left| \Phi_i \right\rangle$ must be chosen adequately and $M$ is the
 basis set size. 

The functions $\left| \Phi_i \right\rangle$ are chosen as 
 eigenfunctions of the total angular momentum with zero eigenvalue. 
The radial part of the basis functions are constructed by symmetrization of the
one-body Slater functions

\begin{equation}\label{slater-type}
\phi^{(\alpha)}_{n}({r}) = \left[ \frac{\alpha^{2n+3}}{(2n+2)!}\right]^{1/2} r^n
e^{-\alpha r/2} .
\end{equation}

With
these constraints the functions $\left| \Phi_i \right\rangle$ have only three
relevant quantum numbers $n_1,n_2$ and $l$, where $n_1$ and $n_2$ are related
to the radial part of the function, and $l$ to the angular one \cite{pont2010}. 
The basis set is characterized by  a non-linear parameter $\alpha$,
then we use  the notation  $\left| \Phi_i \right\rangle \,\rightarrow\,
\left| n_1, n_2;l;\alpha\right\rangle$.
So, we get that
\begin{equation}\label{variational-eigen}
\left|\Psi^{(\alpha)}_j(1,2)\right\rangle = \sum_{n_1 n_2 l}
c^{(i),(\alpha)}_{n_1 n_2 l}
\left|
n_1, n_2;l;\alpha\right\rangle \, ,
\end{equation}
\noindent where $n_1\geq n_2\geq l \ge 0$, then the basis set size is given by
\begin{equation}
M = \sum_{n_1=0}^N \sum_{n_2=0}^{n_1} \sum_{l=0}^{n_2}  1 \,=\,
\frac{1}{6} (N+1) (N+2) (N+3)\; ,
\end{equation}
so we refer to the basis set size using both $N$ and $M$.  
The matrix elements of the kinetic energy, the Coulombic repulsion between the
electrons and other mathematical details involving the functions
$ \left| n_1, n_2 ;l;\alpha\right\rangle$ are given in
references~\cite{pont2010},\cite{osenda2008b},
\cite{pablo-variational-approach}.

\begin{figure}[ht]
\begin{center}
\psfig{figure=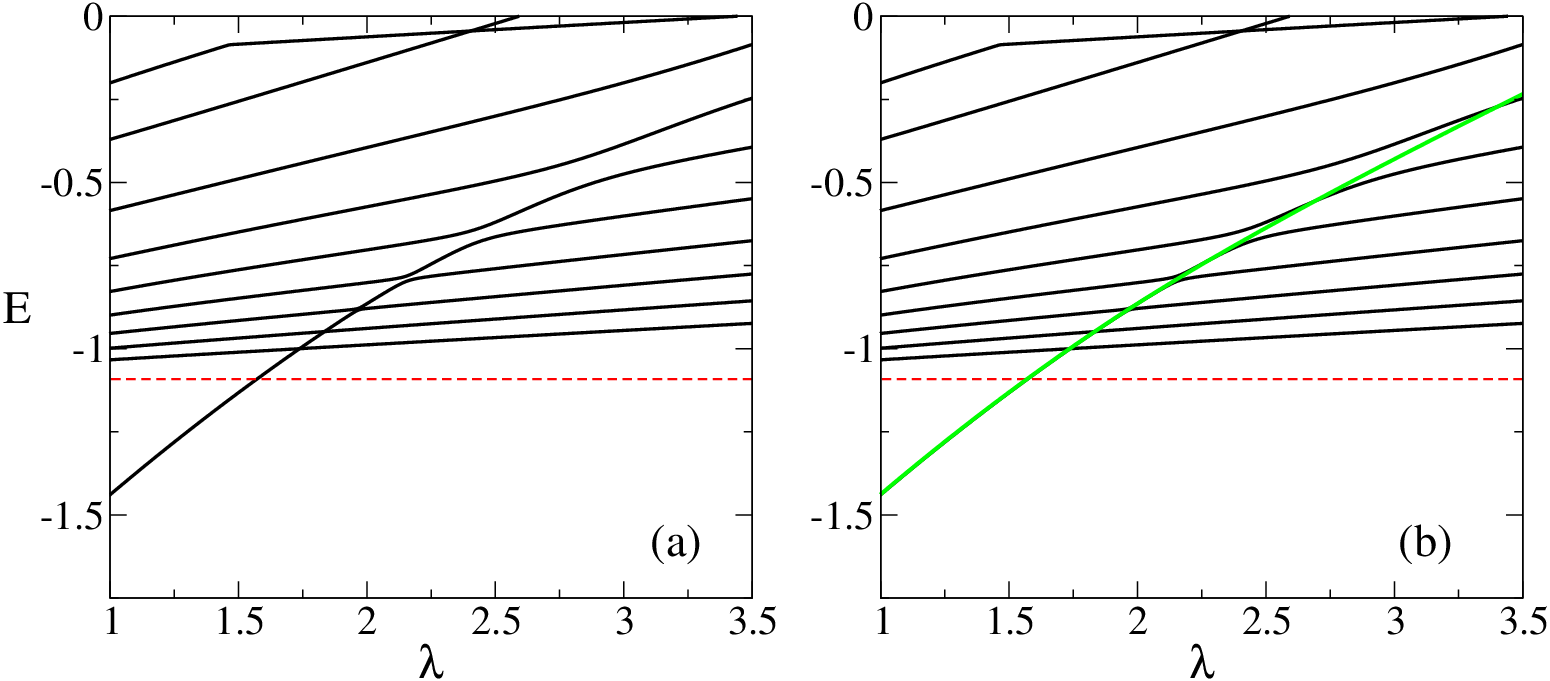,width=14cm}
\end{center}
\caption{\label{avoidedcross}(color on-line) (a) the figure shows the behavior
of the
variational eigenvalues $E_j^{(\alpha)}(\lambda)$ (black lines) for $N=14$ and non-linear parameter $\alpha=2$. The red dashed line
corresponds to the threshold energy $\varepsilon\simeq -1.091$.
Note that the avoided crossings between the variational eigenvalues
are fairly visible. (b) The figure shows the same variational eigenvalues that
(a) (black lines) and the energy calculated using the complex scaling method
(green line) for a parameter $\theta=\pi/10$. }
\end{figure}

Resonance states have isolated complex eigenvalues, $E_{res}=E_r -i \Gamma/2,\;
 \Gamma > 0$, whose eigenfunctions are not square-integrable.
These states are considered as quasi-bound states  of
energy $E_r$ and inverse life time  $\Gamma$. For the Hamiltonian Eq.
(\ref{hamil}), the resonance energies belong to the interval 
$(\varepsilon,0)$ \cite{reinhardt1996}.

The resonance states can be analyzed using the
spectrum obtained with a basis of ${\cal L}^2$ functions (see \cite{ferron2009} and
References therein). 
The levels above the threshold have several avoided crossings 
that ``surround'' the real part of the
energy of the resonance state.
The presence of a resonance
can be made evident looking at the eigenvalues
obtained numerically. Figure~\ref{avoidedcross} shows  a typical
spectrum obtained from the variational method. This
figure shows  the behavior of the variational eigenvalues
$E_j^{(\alpha)}$ as functions of the parameter $\lambda$, the results  were
obtained using $N=14$ and $\alpha=2.0$. The value of $\alpha$ was chosen in
order to obtain the best approximation for the energy of the ground state in
the region of $\lambda$ where it exists. The figure shows
clearly that for
$\lambda<\lambda_{th}\simeq1.54$ there is only one bound state. Above the
threshold there is not a
clear cut criteria to choose the value of the non-linear parameter, and
the variational approximation provides a finite number of solutions
with energy below zero.  However,
it is possible to calculate $E_r(\lambda)$  calculating $E_j^{(\alpha)}$ for many
different values of the variational parameter (see Kar and Ho \cite{kar2004}).

\begin{figure}[ht]
\begin{center}
\psfig{figure=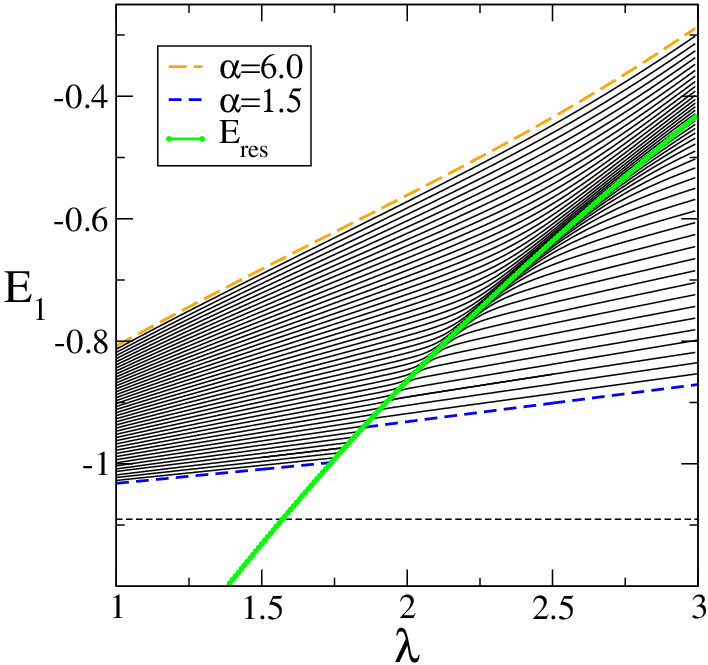,width=8cm}
\end{center}
\caption{\label{densidad-0-1}(color on-line)  The second variational
state energy {\em vs} $\lambda$, for different values of the variational
parameter $\alpha$. From bottom to top $\alpha$ increases its value from
$\alpha=2$ (dashed blue line) to $\alpha=6$ (dashed orange  line). The real part
of the resonance eigenvalue obtained using complex
scaling ( $\theta=\pi/10$) is also shown (green line). }
\end{figure}

Figure~\ref{densidad-0-1} shows the numerical results for the
second
eigenvalue $E_2^{(\alpha)}(\lambda)$, for different values of the variational
parameter
$\alpha$. 
As can be seen from  Figure  ~\ref{densidad-0-1},  there
are three regions, in each
one of them the
curve for a given value of $\alpha$ is basically a straight line and the slope
is different in each region. A feature that appears rather clearly is that, for
fixed $\lambda$ , the
density of levels per unit energy  is not uniform,
despite that the
curves $E_j^{(\alpha_i)}(\lambda)$ are drawn for forty equally spaced
$\alpha_i$'s between $\alpha=2.0$ and $\alpha=6.0$. This
fact has been observed previously \cite{mandelshtam1993} and  the DOS can be
written in terms of two contributions, a localized one and an extended one. The
localized DOS is attributed to the presence of the resonance
state, conversely the extended DOS is attributed to the continuum
of states between $(\varepsilon, 0)$.

\section{The density of states}
\label{sec-density}

The localized DOS $\rho(E)$ can be expressed as \cite{kar2004,mandelshtam1993}
\begin{equation}\label{densidad_sin_suma}
\rho(E)  = \left|\frac{\partial
E(\alpha)}{\partial \alpha}\right|^{-1} . 
\end{equation}
Since we are dealing with a variational approximation, we calculate
\begin{equation}\label{densidad_cal}
\rho(E_j^{(\alpha_i)}(\lambda))  = \left| 
\frac{E_j^{(\alpha_{i+1})}(\lambda) -
E_j^{(\alpha_{i-1})}(\lambda)}{\alpha_{i+1} - \alpha_{i-1}}\right|^{-1} .
\end{equation}
In complex scaling methods the Hamiltonian is dilated by a complex factor 
$\eta=\tilde{\alpha}\,e^{-i\theta}$. As was pointed out long time ago by
Moiseyev and coworkers, the
parameter $\tilde{\alpha}$ is equivalent to our parameter
$\alpha$~\cite{moiseyev79}. Besides, the DOS attains its maximum at
optimal values of $\alpha$ and $E_r$ that could be obtained with a
self-adjoint Hamiltonian without using complex scaling
methods~\cite{moiseyev1980}.
So, locating the position of the resonance using the maximum of the DOS is
equivalent to the stabilization criterium used in complex dilation methods that
requires the approximate fulfillment of the complex virial
theorem \cite{moiseyev1981}.

\begin{figure}[ht]
\begin{center}
  \includegraphics[width=8cm]{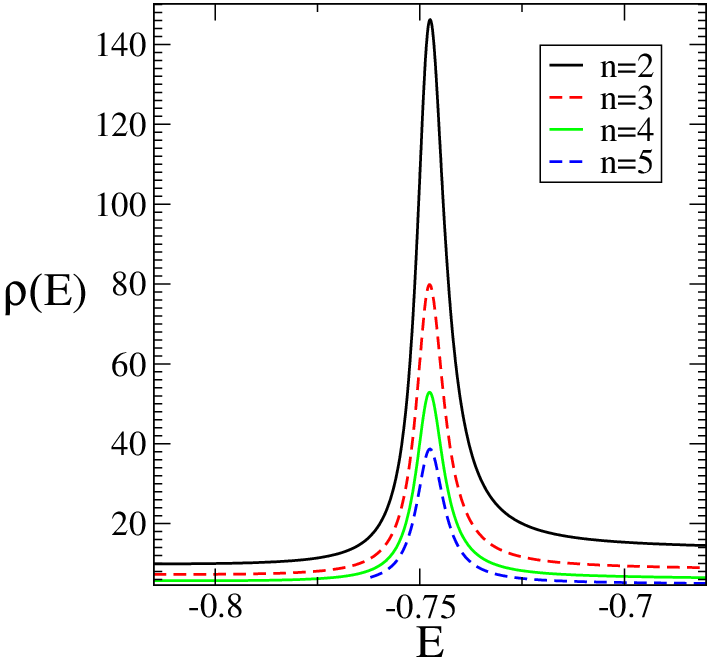}
\caption{\label{fig3_pont} (color on-line) The DOS
$\rho(E)$ for  $\lambda=2.25$ and basis set size $N=14$. The results were
obtained using Eq. (\ref{densidad_sin_suma}) and correspond, from top to
bottom, to the second (black line), third (dashed red line), fourth (green line)
 and fifth (dashed blue line) levels.}
\end{center}
\end{figure}

Figure~\ref{fig3_pont} shows the typical behavior of $\rho_j(E)\equiv
\rho(E_j^{(\alpha_i)}(\lambda))$ for several
eigenenergies, with $\lambda=2.25$. The real
and imaginary parts of the
resonance's
energy, $E_r(\lambda)$ and $E_i(\lambda) =-\Gamma/2$ respectively, 
can be obtained from
$\rho(E)$, see for example \cite{kar2004} and references therein.

The values of $E_r(\lambda)$ and $\Gamma(\lambda)$ are obtained performing
a nonlinear fitting of $\rho(E)${, with a Lorentzian function,}

\begin{equation}
\rho(E)=\rho_0 + \frac{A}{\pi}\frac{\Gamma/2}{\left[(E-E_r)^2
+(\Gamma/2)^2\right]}.
\end{equation}

One of the drawbacks of this method results
evident: for each $\lambda$ there are several $\rho_j(E)$ (in fact one for each
variational level, see Figure~\ref{fig3_pont}), and since each $\rho_j(E)$ 
provides a value for
$E^j_r(\lambda)$ and $\Gamma^j(\lambda)$ one has to choose which one is the
best.
Kar and Ho \cite{kar2004} solve this problem fitting all the $\rho_j(E)$ and
keeping as the best values for $E_r(\lambda)$ and $\Gamma(\lambda)$ the fitting
parameters with the smaller $\chi^2$ value. At least for their data the best
fitting (the smaller $\chi^2$) usually corresponds to the larger $n$. 
This fact has a clear interpretation, if the numerical method approximates
$E_r(\lambda)$ with $E^{(\alpha)}_n(\lambda)$ a large $n$ means that the numerical
method is able to provide a large number of approximate levels, and so the
continuum of states between $(\varepsilon,0)$ is ``better'' approximated. 

In most numerical studies it is assumed that using the
largest possible basis set size leads to the best results. Indeed, most 
usually this
assumption is correct but taking in consideration only the results obtained for
the largest
basis set size does not provide any insight about the accuracy of the result. 

The Finite Size Scaling method (FSS)  provides a systematic way to analyze the
data generated using different basis set sizes and has been extensively used in
atomic physics \cite{ks2003}, statistical mechanics \cite{fisher}, etc. 
The FSS method requires a scaling
function that reduces the data corresponding to different basis set sizes to a
single ``universal'' set. Moreover, as we will show, the FSS allows to obtain
the real and imaginary parts of the resonance energy using any $\rho_j(E)$.

\begin{figure}[ht]
\begin{center}
  \includegraphics[width=14cm]{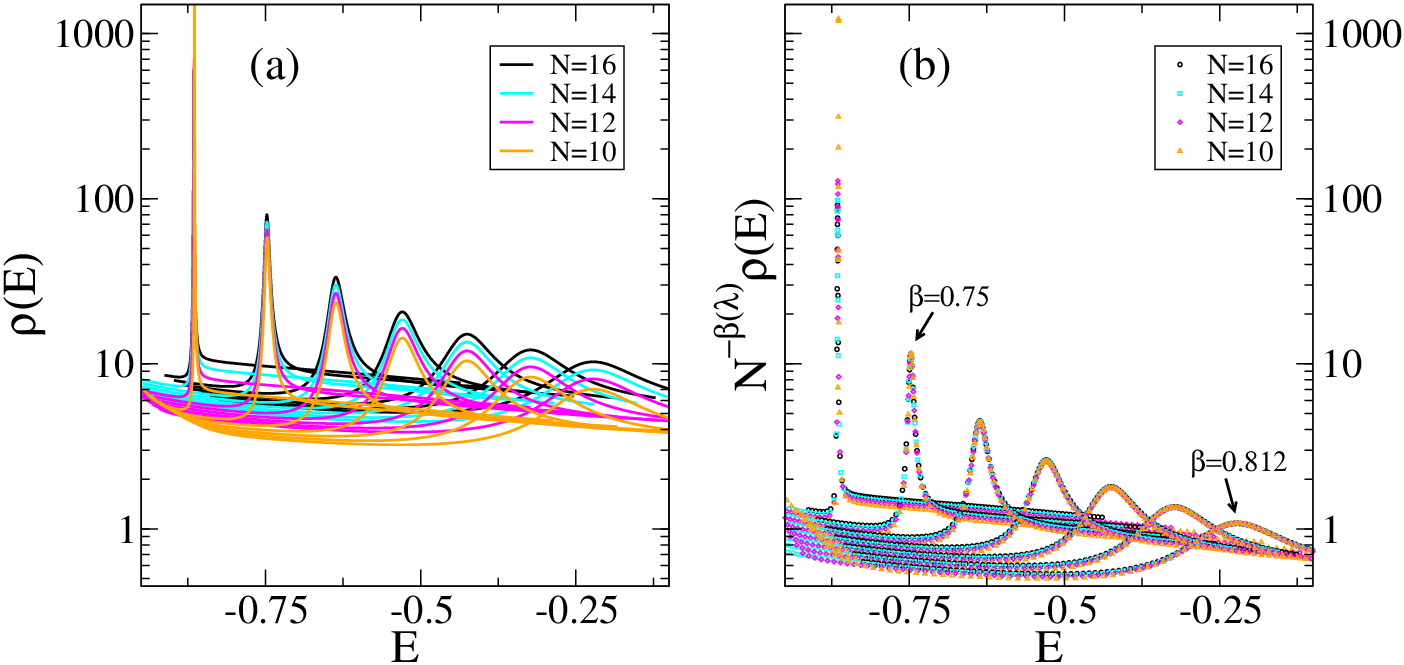}
\caption{\label{densidad} (color on-line)(a) The DOS
$\rho_3(E)$ for  (from left to right) $\lambda=2.25, 2.50,2.75,3.00,3.25$ and
$3.50$ and different basis set sizes. The
results were
obtained using Eq. (\ref{densidad_sin_suma}). (b) Data
collapse obtained using the scaling function $N^{-\beta(\lambda)} \rho(E)$}
\end{center}
\end{figure}

We chose $\rho_3(E)$ to apply the FSS method, but using any other $\rho_j$
should lead to the same conclusions. As we have said previously the
functional form of the scaling function is unique, independently of the actual
value of $\lambda$.
Figure~\ref{densidad}(a) shows $\rho_3(E)$ for different values of $\lambda$
and several basis set sizes. For a fixed value of $N$, say $N=16$ the
Figure ~\ref{densidad}(a) shows seven peaks, each one corresponding to a given
value of $\lambda$. The values of $\lambda$ were chosen to sample different
regimes of the resonance, from small values -sharp resonances- to larger ones
-broad resonances. The broadening of the peaks corresponds roughly to the
broadening of the resonance or, equivalently, to larger values of $\Gamma$.

Figure~\ref{densidad} (b) shows the scaled data  of panel (a), for each
$\lambda$ the curves corresponding to different values of $N$ collapse into a
single one which is independent of $N$. 

\begin{figure}[ht]
\begin{center}
  \includegraphics[width=14cm]{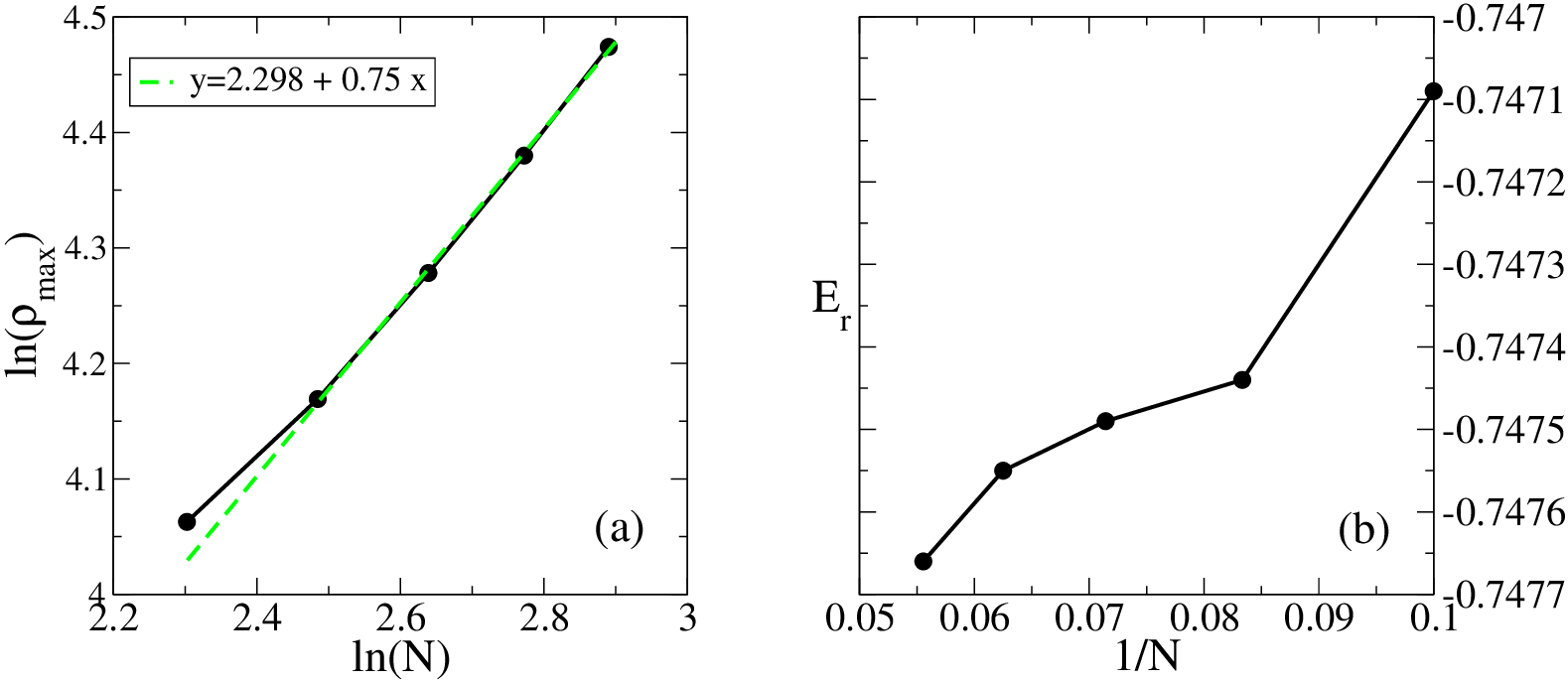}
\caption{\label{scaling-function} (color on-line)  
(a) $log-log$ plot of the  maximum of the density of states against the
basis-set size $N$. (b) $E_r$ vs. $1/N$. For both panels the data were obtained 
with  $\lambda=2{.}25$.}
\end{center}
\end{figure}

The form of the scaling function can be inferred from the
behavior of the maximum value of the function $\rho_3(E)$ for different basis
set sizes. Figure~\ref{scaling-function} shows the behavior of
$\ln({\rho_3}_{max})$ {\em vs} $\ln(N)$,  for $N$ large enough the
maximum values behave as a power law. With this numerical evidence we propose
\begin{equation}
 \rho(E) \longmapsto N^{-\beta(\lambda)} \rho(E)
\end{equation}
as the scaling function for the DOS. As can be seen in
Figure~\ref{densidad} (b), the collapse of the data is excellent. The exponent
$\beta$ depends on $\lambda$ and its value must be found for each case. So, for
a given $\lambda$ the real part of the complex resonance energy corresponds to
the value where the DOS attains its maximum, $E_r(\lambda) =
E_{\rho_{max}}$, where $\rho_{max} = \max_E \rho(E)=\rho(E_{\rho_{max}})$. At
least from our data it is clear that the localization  of the maximum of the 
DOS is a
very slowly function of the basis set size, see Figure~\ref{scaling-function}
(b). 

\begin{figure}[ht]
\begin{center}
  \includegraphics[width=14cm]{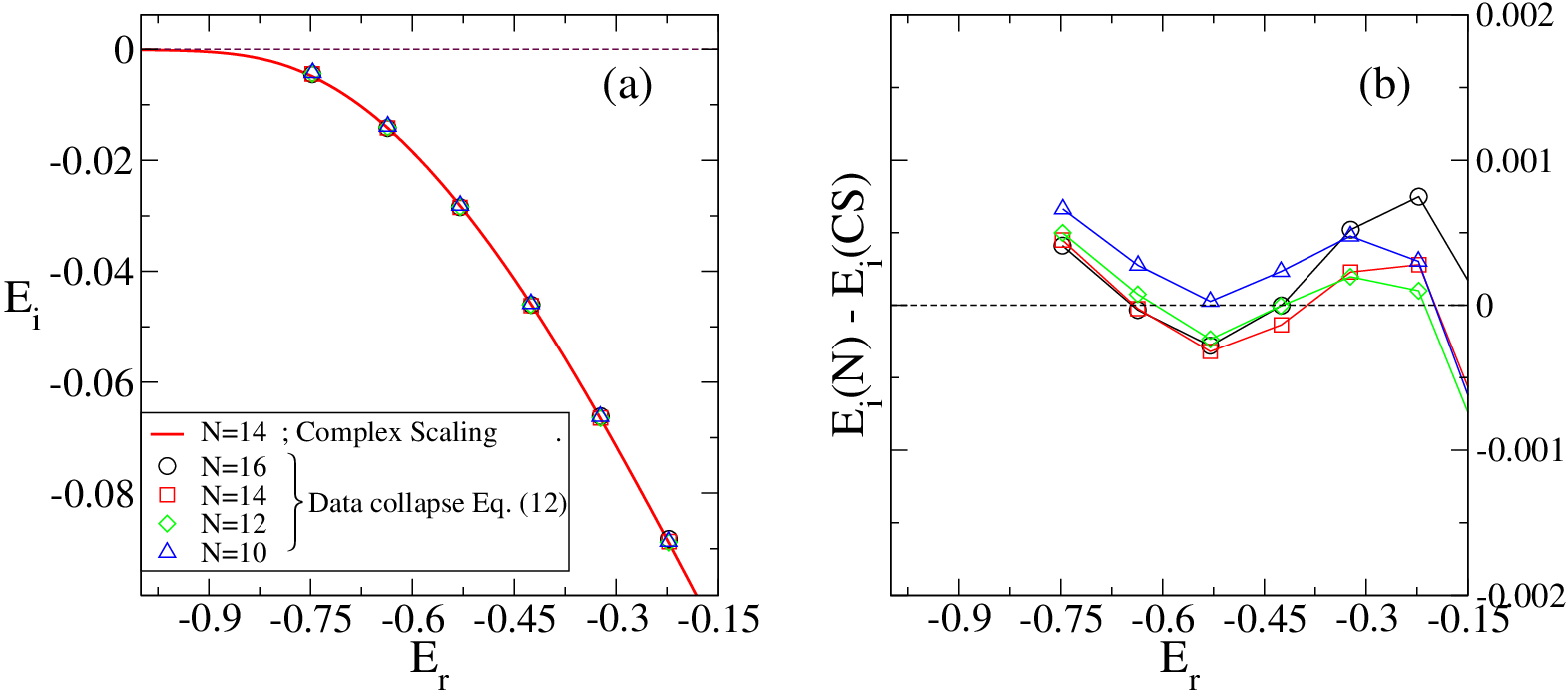}
\caption{\label{real-imag} (color on-line) (a) Resonance pole string curve obtained
from complex scaling and basis-set size scaling. A mean value of $\delta=0.8$ was
chosen for the scaling exponent. (b) The energy discrepancies between the two methods.
The relative error is less than $1\%$ for the calculated values.}
\end{center}
\end{figure}

On the following we will argue about the form of the scaling function necessary
to obtain the imaginary part of the resonance energy, $\Gamma/2$. As we have
said previously, the real part of the resonance energy is located
between $(\varepsilon,0)$, and as log as $\Gamma \rightarrow 0$ when $\lambda
\rightarrow \lambda_c^+$, the scaling function proposed for the imaginary part
of the resonance energy should be zero at $\lambda_c$ or, equivalently,
$E_r(\lambda_c)=\varepsilon$, and  $E_i(\lambda_c) = -\Gamma(\lambda_c)/2=0$.
Besides, in some examples the life time of a resonance is proportional to the
DOS, $\Gamma \propto \frac{1}{\rho}$, so we tested our numerical
results using 
\begin{equation}\label{gamma-scaled}
E^{(N)}_i(\lambda) = - \kappa   \; (E^{(N)}_r(\lambda) - E_c) N^{\delta}
\frac{1}{\rho^{(N)}_{max}},
\end{equation}
where we have made explicit the superscript $N$ to emphasize the dependency
with the basis set size,  kappa is a positive constant, and  $\delta$ is a 
exponent to be
determined. As a first guess we used $\delta=0.8$, since this figure is very
close to the exponents founded studying the scaling of the maximum of the
DOS. Then, we observed that 
\begin{equation}
\frac{ E^{(N)}_i(\lambda^{\prime \prime})}{E^{(N)}_i(\lambda^{\prime})} \simeq
\frac{E_i^{(CS)}(\lambda^{\prime \prime})}{ E^{(CS)}_i(\lambda^{\prime})},
\end{equation}
{\em i.e.} the ratio between the imaginary parts of two resonance
energies, corresponding to $\lambda^{\prime \prime}$ and $\lambda^{\prime}$
chosen arbitrarily, calculated using the Complex Scaling (CS) method and
Eq.~(\ref{gamma-scaled}) {\bf is fairly the same}. Regrettably this finding
does not provide the value of the constant $\kappa$. Just to compare with the
result of Complex Scaling methods \cite{pont2010}, we calculated $\kappa$ 
imposing
that $ E^{(N)}_i(\lambda) = E^{(CS)}_i(\lambda)$ for one particular $\lambda$. 
Figure~\ref{real-imag} shows the values obtained for $E_i^{(N)}$ obtained using
the recipe described above and Eq.~(\ref{gamma-scaled}), the agreement
between values corresponding to different basis set sizes is excellent, and
the same can be said with respect to the values obtained from Complex Scaling,
see Figure~\ref{real-imag} (a). Despite that the value of $\kappa$ was obtained
using a particular value of $N$ and $\lambda$ our numerical findings
strongly support the functional form proposed in Eq.~(\ref{gamma-scaled}),
in Figure~\ref{real-imag} (b) we present a comparison between the
$E^{(N)}_i(\lambda)$'s and the Complex Scaling results, the agreement is better
than $1\%$ for all the values of $\lambda$ and $N$ analyzed.

\section{Summary and conclusions}
\label{sec-conclusions}

So far we have presented numerical evidence of the critical behavior of the
density of states near the peak associated to the presence of a resonance
state. The critical behavior should be understood in the sense of FSS method,
{\em i.e} there is a strong dependency of the density of states with the basis
set size used to obtain it. The critical behavior also manifest itself through
the scaling properties of some quantities, in particular the maximum value
attained by the density of states for fixed $\lambda$ and $N$.

Despite that we have been able to propose a scaling function that provides the
imaginary part of the resonance energy, the proposal is made only in
phenomenological terms and, more importantly, relays in other methods to be
quantitative. Our efforts will be directed to put the scaling law
 Eq.(\ref{gamma-scaled}) over a more formal basis, which
hopefully will allow us to obtain the constants $\kappa$ and $\delta$ using only
scaling arguments. 

The critical behavior of the density of states studied in this work, together
with the behavior observed in the fidelity \cite{pont2010}, points that the
analogy between the ``line'' of resonance states and a ``critical line''
(in the sense of critical phenomena) could be more exploited and it is worth of
further exploration.

\ack
We would like to acknowledge  SECYT-UNC,  CONICET and FONCyT 
for partial financial support of this project.

\end{document}